\documentclass[prb,aps,10pt,twocolumn, superscriptaddress, showpacs,floatfix,letterpaper]{revtex4-1}  
\usepackage{hyperref}
\usepackage{amssymb}
\usepackage{amsmath}
\usepackage{float}
\usepackage{bbm}
\usepackage{graphicx}
\usepackage{epsfig}
\usepackage{epstopdf}
\usepackage[usenames]{color}
\usepackage{overpic}

\begin{document}
\bibliographystyle{apsrev4-1}

\title{Quasiparticle  Properties of the Superconducting State of the Two Dimensional Hubbard Model}

\author{E. Gull}
\affiliation{Department of Physics, University of Michigan, Ann Arbor, Michigan 48109, USA}

\author{A. J. Millis }
\affiliation{Department of Physics, Columbia University, New York, New  York 10027, USA}

\date{\today }

\begin{abstract}
Cluster dynamical mean field methods are used to calculate the normal and anomalous components of the electron self energy of the two dimensional Hubbard model. Issues associated with the analytical continuation of the normal and anomalous parts of the gap function are discussed. Methods of minimizing the uncertainties associated with the pseudogap-related pole in the self energy are discussed. From these  the evolution of the superconducting gap and the momentum dependent electron spectral function across the phase diagram are determined.  In the pseudogap regime, decreasing the temperature into the superconducting state leads to a decrease in the energy gap and the formation of a `peak-dip-hump' structure in the electronic density of states. The peak feature disperses very weakly. The calculated spectral functions are in good qualitative agreement with published data.   The mathematical origin of the behavior is found to be the effect of the superconductivity on the pole structure giving rise to the normal state pseudogap. In particular the ``hump'' feature is found to arise from a zero crossing of the real part of the electron self energy rather than from an onset of scattering. The effect of superconductivity on the zone diagonal spectra is presented.  \end{abstract}


\maketitle
\section{Overview}
After more than a quarter century of research, our theoretical understanding of the unusual electronic properties of the  high transition temperature superconductivity observed \cite{Bednorz86} in layered copper oxide materials remains incomplete. A plethora of remarkable dynamical phenomena have been reported,  including a correlation-driven insulating phase occurring when the conduction band is half filled,\cite{Anderson87} non-Fermi-liquid transport properties,\cite{Taillefer10} a `pseudogap' in the electronic spectrum \cite{Huefner08} as well as ordered phases including `stripe' states with spin and charge order,\cite{Tranquada95} charge order apparently unconnected with spin order,\cite{Ghiringhelli12} `nematic' (rotational symmetry breaking) states,\cite{Hinkov08} and time reversal symmetry breaking states also apparently unaccompanied by conventional spin order.\cite{Fauque06} Angle-resolved photoemission studies report unusual quasiparticle properties at essentially all carrier concentrations.\cite{Ding96,Norman97,Norman98,Shen98,Valla00,Johnson01,Lanzara01,Kordyuk02,Borisenko03,Koitzsch04,Kaminsky05,Kanigel08,He11} 

The  diversity  of reported phenomena has led to debate  about what is the essential physics to include in a minimal theoretical model while the apparent strong coupling nature of the problem suggests that whatever model is adopted, a nonperturbative treatment is required.   Important open questions include the mechanism for superconductivity, the nature of the minimal low energy model describing the high-$T_c$ phenomenon,\cite{Anderson87,Scalapino94} and the physics of the `pseudogap' and its interplay with the superconducting gap.\cite{He11,Gull13}  

This paper presents a theoretical study of the electron excitation spectrum of the superconducting and normal states of the two dimensional Hubbard model. This model is the minimal model of the physics of strongly correlated electrons on a lattice. Although it is not yet known if this model contains the full panoply of phenomena observed in the high-$T_c$ materials, it clearly contains some important aspects of the physics and is accepted as one of the candidate models \cite{Anderson87} for describing the low energy (energies of order $1eV$ or less) physics of the copper-oxide superconductors. Determining the properties of this model to the level at which a clear comparison to experiment can be made is  an important goal of theory.  

We address this problem using the `dynamical cluster approximation' (DCA) version \cite{Hettler98,Hettler00} of the cluster dynamical mean field method.\cite{Maier05} The method is based on approximating the full spatial dependence of the electron self energy in terms of a finite number $N_c$ of functions of frequency,\cite{Okamoto03} with the exact properties recovered in the $N\rightarrow\infty$ limit.\cite{LeBlanc13} Within this approximation the method provides an unbiased (in the sense of not pre-selecting a particular interaction channel or class of diagrams) numerical approach to the correlated electron problem and allows comprehensive investigation of the frequency dependence, and some aspects of the momentum dependence, of the electronic properties.

Our ability to solve the equations of dynamical mean field theory\cite{Gull08,Gull10_submatrix} has reached  the point where approximation  sizes $N_c$ that in many aspects are representative \cite{Gull10_clustercompare}  of the $N_c\rightarrow\infty$ limit can be studied at the low temperatures required to stabilize superconductivity. \cite{Gull13} The superconducting state has been constructed \cite{Gull13} and physical properties including the superconducting condensation energy,\cite{Gull12b} the  c-axis  and Raman response \cite{Gull13b} and the structure of the gap function and pairing potential \cite{Gull14_glue} have been computed and found to be in remarkable agreement with experiment.  Here we use the new methodologies to study the electronic excitation spectrum of the normal and superconducting states in detail. While some of these issues have been previously studied in a cluster size  $N_c=4$ approximation, \cite{Maier00,Maier05,Civelli09,Civelli09b,Kyung09,Senechal13,Sakai14} our results, obtained on larger $N_c=8$ clusters,  provide a better representation of the physics including a clear separation of nodal and antinodal behavior.

The DCA equations are solved in imaginary time, and an analytical continuation\cite{Jarrell96} procedure is needed to obtain the real frequency information needed for the electronic excitation spectrum. Especially in the pseudogap regime of the model, the novel physics we find presents challenges for the  analytical continuation of our computed results. These are discussed at length below. 

The rest of the paper is organized as follows. In Section~\ref{Methods} we define the quantities of interest and present the specifics of the dynamical mean field method. Section~\ref{continuation} discusses issues related to the analytic continuation.   Section~\ref{self_energies} displays the normal and anomalous components of the self energy, drawing attention to an unusual pole structure related to sector-selective Mott nature of the pseudogap phase.\cite{Ferrero09,Werner098site,Gull09} Section~\ref{gap} displays the gap function constructed from the ratio of anomalous and normal components of the self energy. Sections \ref{spectraFS}  and \ref{spectra} display the photoemission and inverse photoemission spectra predicted by the model while section~\ref{diagonal} presents our findings on superconductivity-induced changes to the electron scattering and mass renormalization for states near the zone diagonal where the superconducting order parameter vanishes. Section~\ref{Summary} presents a summary and conclusions.

\section{Formalism \label{Methods}}
\subsection{Model}
We study the two dimensional Hubbard model of electrons hopping on a square lattice and subject to a local interaction $U$ which we take to be repulsive. The model may be written in a mixed momentum ($k$)/position ($i$) representation as
\begin{align}
H=\sum_{k}\text{Tr}\left[\Psi^\dagger_k\tau_3\left(\varepsilon_k-\mu\right)\Psi_k\right]+U\sum_in_{i\uparrow}n_{i\downarrow}.
\label{Hhub}
\end{align}
In the first term we have represented the electronic degrees of freedom by the Nambu spinor defined in terms of $c^\dagger_{k\sigma}$, the Fourier transform to momentum space of the operator $c^\dagger_{i\sigma}$ which creates an electron of spin $\sigma=\uparrow,\downarrow$ on lattice  site $i$ as
\begin{align}
\Psi^\dagger_k=\left(\begin{array}{cc}c^\dagger_{k\uparrow} &c_{-k\downarrow}\end{array}\right)\ \ ;\ \ 
\Psi_k=\left(\begin{array}{c}c_{k\uparrow} \\c^\dagger_{-k\downarrow}\end{array}\right).
\label{psidef}
\end{align}
We  set the lattice constant to unity. The momentum index $k$ runs over the Brillouin zone of the two dimensional square lattice $-\pi\leq k_x,~k_y\leq\pi$. The trace is over the Nambu indices and $\tau_j$ denote the Pauli matrices operating in Nambu space.  The chemical potential is $\mu$, $\varepsilon_k$ is the energy dispersion and  $n_{i\sigma}=c^\dagger_{i\sigma}c_{i\sigma}$ is the operator measuring the density of spin $\sigma$ electrons on site $i$.   In the computations presented below we take  $\varepsilon_k=-2t\left(\cos~k_x+\cos~k_y\right)$  and present our results in units of $t$. A reasonable estimate of the energy scales pertaining to the physical copper oxide materials is $t\approx 0.3eV$.   At carrier density $n=1$ per site this version of the Hubbard model is particle-hole symmetric ($\mu=0$), but particle-hole symmetry is broken at carrier concentrations $n\neq 1$.

\subsection{Green function and self energy}
Our analysis proceeds from  the components of the matrix Nambu Green function defined for imaginary time $\beta=1/T>\tau>0$ (T is the temperature) as
\begin{align}
\mathbf{G}(k,\tau)=-\left\langle\Psi_k(\tau)\Psi^\dagger_k(0)\right\rangle.
\label{GNambu}
\end{align} 
It is useful to Fourier transform $G(\tau)$ to the Matsubara frequency  axis
\begin{equation}
\mathbf{G}(k,i\omega_n) = \int_0^\beta d\tau \mathbf{G}(k, \tau)e^{i\omega_n\tau}
\label{Gomega}
\end{equation}
with $\omega_n=(2n+1)\pi T$. The self energy is defined as
\begin{align}
\mathbf{\Sigma}(k,i\omega_n)=\begin{pmatrix}i\omega_n-\varepsilon_k+\mu \hspace{-0.5cm}&0 \\0 &\hspace{-0.5cm} i\omega_n+\varepsilon_k-\mu\end{pmatrix}-\mathbf{G}^{-1}(k,i\omega_n)
\label{sigdefgf}
\end{align}
and may be  written  explicitly in terms of normal (N) and anomalous (A) components as (note we have chosen the phase of the superconducting order parameter to be real)
\begin{align}
{\mathbf \Sigma}(k,i\omega_n)=\left(\begin{array}{cc}
\Sigma^N(k,i\omega_n) & \Sigma^A(k,i\omega_n) \\
 \Sigma^A(k,i\omega_n) & -\Sigma^N(k,-i\omega_n)\end{array}\right).
\label{sigdef}
\end{align}
Under spatial (k) transformations $\Sigma^N(k,i\omega_n)$ has the full symmetry of the lattice, while depending on the superconducting state  $\Sigma^A(k,i\omega_n)$ may have lower symmetry. Our investigations,\cite{Gull13} consistent with a large body of previous work, \cite{Zanchi96,Gonzales00,Halboth00,Lichtenstein00,Maier00,Honerkamp01,Kampf03,Maier05_dwave,Haule07,Scalapino07,Civelli08,Maier08,Kancharla08,Raghu10,Sordi12,Sordi13} indicate that for the interesting carrier concentrations $|1-n|<0.3$ and for temperatures higher than $T=t/60$  the only stable superconducting state of this model  is of $d_{x^2-y^2}$ symmetry, so that on the square lattice studied here $\Sigma^A(k,i\omega_n)$ changes sign if $k$ is rotated by $\pi/2$ or reflected through one of the axes $k_y=\pm k_x$ that lie at $45^\circ$ to the lattice vectors, but remains invariant under reflections through the bond axes $k_x=0$ or $k_y=0$  and under rotations by $\pi$. 

We now discuss the frequency dependence at fixed $k$. In this discussion because the wave vector is fixed we do not explicitly denote it. The anomalous self energy is an even function of Matsubara frequency: $\Sigma^A(i\omega_n)=\Sigma^A(-i\omega_n)$. However the normal component at positive Matsubara frequency $\Sigma^N(i\omega_n)$ is not simply related to the normal component at negative Matsubara frequency $\Sigma^N(-i\omega_n)$ except in the special case of particle-hole symmetry where $\Sigma^N(i\omega_n)=-\Sigma^N(-i\omega_n)$. For later purposes it is convenient to distinguish  components of $\Sigma^N(i\omega_n)$ even (e) and odd (o) in Matsubara frequency\cite{Gull14_glue} as
\begin{equation}
\Sigma^N_{e,o}(i\omega_n)=\frac{\Sigma^N(i\omega_n)\pm\Sigma^N(-i\omega_n)}{2}
\label{SigmaNevenodd}
\end{equation}
so that in Nambu notation we may write
\begin{eqnarray}
&&\mathbf{\Sigma}(i\omega_n)=\Sigma^N_o(i\omega_n)\tau_0+\Sigma^N_e(i\omega_n)\mathbf{\tau}_3+\Sigma^A(i\omega_n)\tau_1.
\label{SigmaNambu}
\end{eqnarray}

It is convenient also to define the gap function $\Delta(i\omega_n)$ as\cite{Poilblanc02,Gull14_glue}
\begin{equation}
\Delta(i\omega_n)=\frac{\Sigma^A(i\omega_n)}{1-\frac{\Sigma_o(i\omega_n)}{i\omega_n}}.
\label{Deltadef}
\end{equation}
This definition is motivated by the observation that at the low frequencies of interest here, the even (under sign change of Matsubara frequency) component of $\Sigma^N$ can be absorbed into a shift of chemical potential. 

\subsection{Analytic Structure}
$\mathbf{G}(z)$ and $\mathbf{\Sigma}(z)$ are analytic functions of complex frequency argument $z$ except for a branch cut along the real axis $\text{Im}(z)=0$. Apart from a Hartree term in the normal component of $\Sigma$ they decay as $\left|z\right|\rightarrow\infty$. They thus obey a Kramers-Kronig relation (written here for $\mathbf{\Sigma}$, with $\mathbf{\Sigma}^{(2)}$ one half of  the branch-cut discontinuity)
\begin{equation}
\mathbf{\Sigma}(z)=\int \frac{dx}{\pi}\frac{\mathbf{\Sigma}^{(2)}(x)}{z-x}.
\label{KKSigma}
\end{equation}
$\mathbf{\Sigma}^{(2)}(\omega)$ is a matrix with  eigenvalues that are non-negative and related by time reversal, thus:
\begin{equation}
\mathbf{\Sigma}^{(2)}(\omega)=\mathbf{R}^\dagger_\omega\left(\begin{array}{cc}s(\omega) & 0 \\0 & s(-\omega)\end{array}\right)\mathbf{R}_\omega
\label{spectral}
\end{equation}
with $s(\omega) \geq 0$. Here
\begin{equation}
\mathbf{R}_\omega=\exp\left[\frac{i}{2}\theta_\omega\tau_2\right]
\label{Rdef}
\end{equation}
is a rotation matrix in Nambu space parametrized by an angle  $\theta_\omega$  which in general is frequency dependent and lies in the range $0\leq\theta_\omega=\theta_{-\omega}\leq\frac{\pi}{2}$. That only $\tau_2$  appears in Eq.~\ref{Rdef} follows from our phase convention for the superconducting state, Eq.~\ref{SigmaNambu}.

Evaluating Eq.~\ref{spectral} for $z$ approaching the real axis gives the real-time Green functions. Of particular interest are the retarded functions obtained by continuation from below to the real frequency axis: $z\rightarrow \omega-i\delta$ with $\omega$ real and  $\delta$ (typically not explicitly written) a positive infinitesimal.  The retarded real axis axis functions have real and imaginary parts $\mathbf{G}=\text{Re}\left[\mathbf{G}\right]+i\text{Im}\left[\mathbf{G}\right]$ and $\mathbf{\Sigma}=\text{Re}\left[\mathbf{\Sigma}\right]+i\text{Im}\left[\mathbf{\Sigma}\right]$, respectively.  Our choice of superconducting phase convention (upper right and lower left entries in self energy matrix identical) implies that $\mathbf{R}$ is purely real so that $\text{Im}\mathbf{\Sigma}(z\rightarrow \omega-i\delta)$ is identical to the branch cut discontinuity $\mathbf{\Sigma}^{(2)}$ introduced in Eq.~\ref{KKSigma}.

The Kramers-Kronig relation  along with the positivity of $\text{Im}\Sigma^N$ implies that $\Sigma^N_e(i\omega_n)$ and $\Sigma^A(i\omega_n)$ are a real functions of $\omega_n$ while $\Sigma^N_o(i\omega_n)$ is a purely imaginary  function of $\omega_n$.
The symmetry properties of $\Delta(\omega)$ are those of $\Sigma^A(\omega)$. 

The  non-negativity of the two eigenvalues of the spectral function implies that the diagonal components of the imaginary parts of $\mathbf{\Sigma}$ and $\mathbf{G}$ are non-negative so that on the real frequency axis $\text{Im}\Sigma_o^N(\omega)>0$ and  $\left|\text{Im}~\Sigma^N_o(\omega)\right|\geq \left|\text{Im}~\Sigma^N_e(\omega)\right|$. 

\subsection{Method}
To solve the Hubbard model we employ the dynamical mean field approximation \cite{Metzner89,Georges92,Georges96} in its DCA cluster\cite{Hettler98,Maier05} form. In this approximation the Brillouin zone is partitioned into $N_c$ equal area tiles labeled by central momentum $K$ and  the self energy is approximated as a piecewise continuous function
\begin{align}
\mathbf{\Sigma}(k,\omega)\approx\sum_K^{N_c}\phi_K(k)\mathbf{\Sigma}_K(\omega)
\label{DCA}
\end{align}
with $\phi_K(k)=1$ if $k$ is in the tile centered on $K$ and 0 otherwise. The functions $\mathbf{\Sigma}_K(\omega)$ are Nambu matrices with normal (N) and anomalous (A) components which are determined from the solution of an auxiliary quantum impurity model as described in detail in Refs.~\onlinecite{Lichtenstein00,Maier06,Gull11,Gull13}.

The quantum impurity model is solved by the continuous-time auxiliary field quantum Monte-Carlo method introduced in Ref.~\onlinecite{Gull08} and discussed in detail in Ref.~\onlinecite{Gull11}. Fast update techniques \cite{Gull10_submatrix} are crucial for accessing the range of interaction strengths and  temperatures needed to study superconductivity and its interplay with the pseudogap. The quantum Monte-Carlo calculations are performed on the Matsubara axis and maximum-entropy analytical continuation techniques\cite{Jarrell96,Wang09a} are employed to obtain real-frequency results.   

\begin{figure}[t]
	\includegraphics[width=0.38\columnwidth]{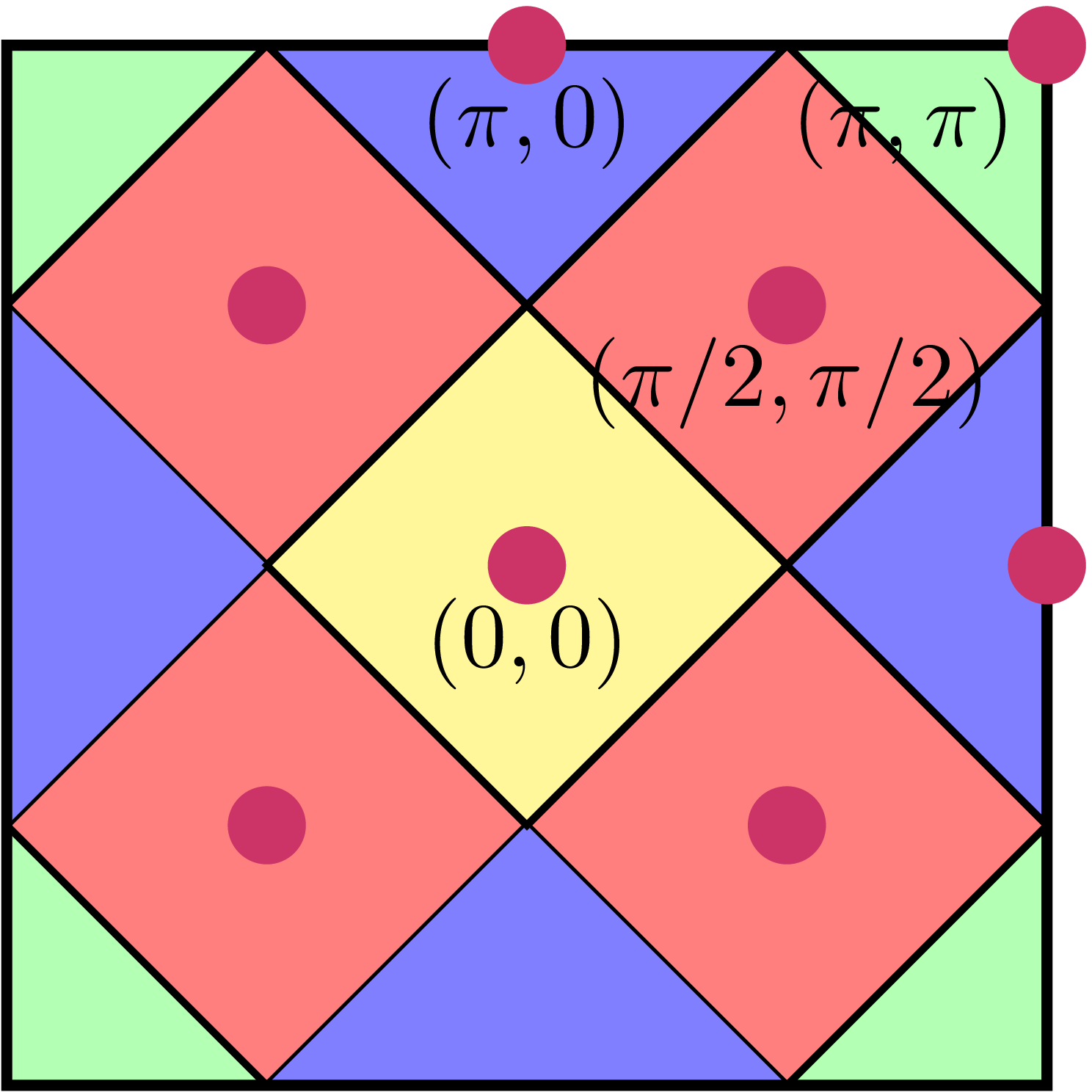}
	\includegraphics[width=0.60\columnwidth]{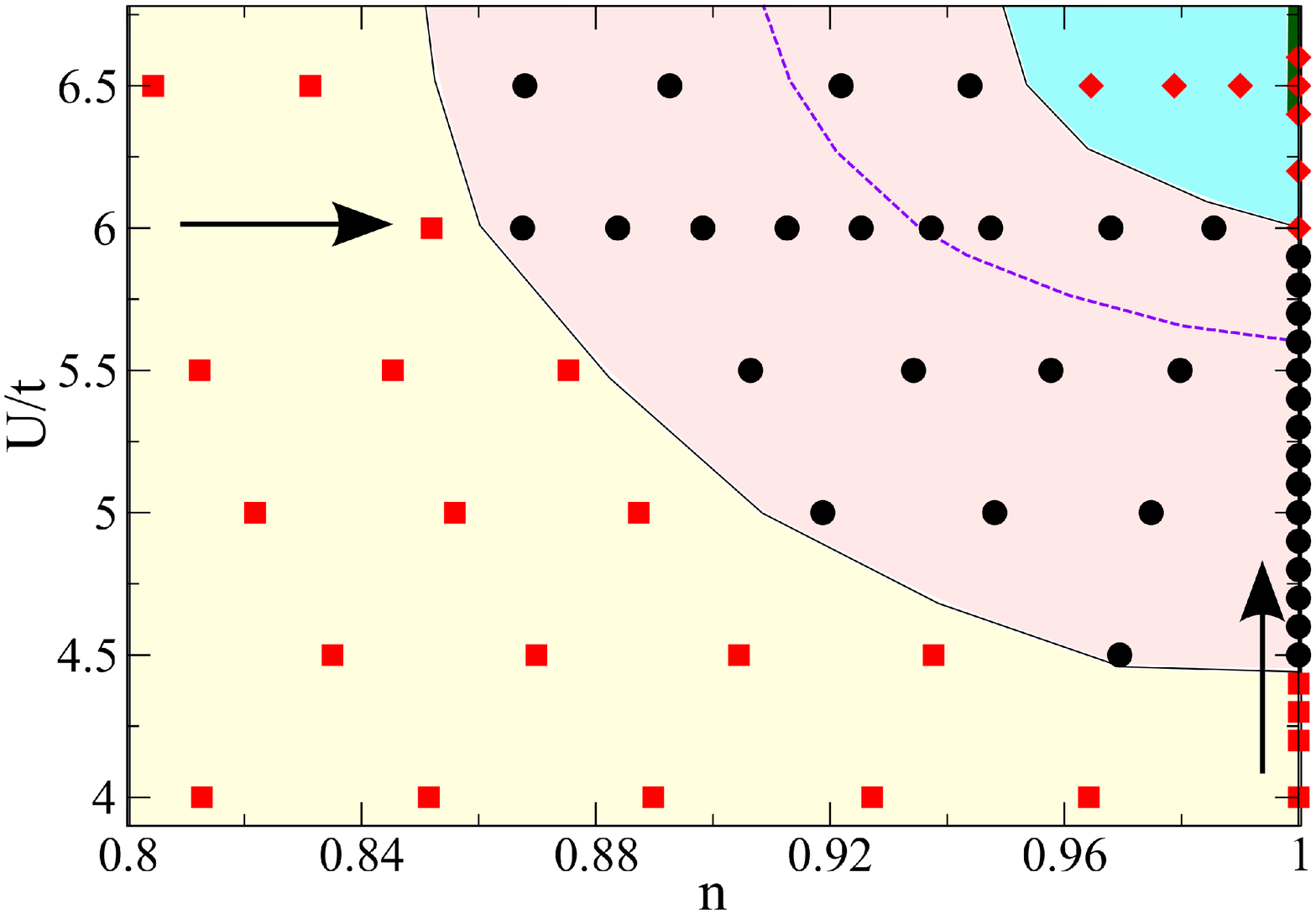}
	\caption{Left panel: Momentum-space tiling used in the present study. Different colors represent patches on which the self-energy is constant.  Right panel: Phase diagram in space of interaction strength and carrier concentration.  A Mott insulating phase (heavy line, green online)  is found at density $n=1$ and interaction strength $U\geq6.5t$.   A non-superconducting, pseudo gapped phase (diamonds and  darker shading, blue online) separates the Mott insulator from  a superconducting phase of $d_{x^2-y^2}$ symmetry  (circles, intermediate shading, pink online). At larger doping or weaker correlation strength a non-superconducting (at the temperatures accessible to us) Fermi liquid phase is found (squares, lightest shading, yellow online).  The onset of the normal state pseudogap is indicated by a light dotted line (purple online) running through the superconducting phase.   In the pseudogapped regime the electronic spectrum is gapped for momenta near the zone face but is gapless for momenta near the zone diagonal.. Figure reproduced from Ref.~\onlinecite{Gull12b}\label{fig:pdtiling}.}
\end{figure}

The expense of the computation increases rapidly as the  interaction strength $U$ or  number of approximants $N_c$  increases, or as the temperature $T$ decreases. We present results for $N_c=8$ using the momentum-space tiling shown in the left panel of Fig.~\ref{fig:pdtiling}. Previous work \cite{Gull10_clustercompare,Gull13} has shown that this cluster is in many aspects representative of the $N_c\rightarrow\infty$ limit; in particular it is large enough to enable a clear distinction between zone-diagonal and zone-face electronic properties, yet small enough to permit  calculations in the superconducting phase of the precision needed for analytical continuation. 

The d-wave symmetry of the superconducting state of the two dimensional Hubbard model means that  in the coarse-grained momentum resolution available in the $N_c=8$ DCA we have 
\begin{align}
\Sigma^A(k,\omega) =
\left\{
\begin{array}{r l}
\Sigma^A(\omega), & k\in (\pi,0)\\
0, & k\in (\pm\frac{\pi}{2},\pm\frac{\pi}{2}), ~(0,0)~(\pi,\pi)\\
-\Sigma^A(\omega), &k\in (0,\pi).
\end{array}
\right.
\end{align}
$\Sigma^N$  has the full point group symmetry of the lattice and thus  is one function of frequency in the $(\pi,0)$ and $(0,\pi)$ sectors, a different function of frequency in the $(\pm \frac{\pi}{2},\pm\frac{\pi}{2})$ sectors and yet a different function of frequency in the $(0,0)$ and in the $(\pi,\pi)$ sectors. The main focus of attention in this paper will be on the $K=(\pi,0)$  momentum sector but some results will be presented on superconductivity-induced changes in the zone-diagonal $(\pm \frac{\pi}{2},\pm\frac{\pi}{2})$ momentum sectors. Because in the 8-site DCA the anomalous self energy is non-zero only in the $K=(0,\pi)/(\pi,0)$ sectors we will typically omit the momentum argument in our discussions of $\Sigma^A$ and $\Delta$. 

The right panel of Fig.~\ref{fig:pdtiling} shows the phase diagram in the plane of interaction strength and carrier concentration obtained \cite{Gull12b,Gull13} from the solution of the DCA equations at temperature $T=t/60$, about half of the maximal computed superconducting transition temperature $T_c^\text{max}\approx t/30$. The temperature $T=t/60$ corresponds in physical units to   $T\approx 60K$.\cite{Gull09,Gull10_clustercompare}

We study the electronic properties as a function of doping at $U=6t$ and as a function of interaction strength at carrier concentration $n=1$.  $U=6t$ is the largest interaction strength for which high precision data could be obtained with the resources available to us for temperatures substantially below $T_c$ and general dopings. Simulations at larger interaction strengths are severely hampered by the fermionic sign problem. As can be seen from the phase diagram, this interaction strength is such that the model is conducting (although pseudo gapped) at $n=1$. Thus, we believe that $U=6t$ is slightly lower than the $U$ which is relevant to the real materials, so the quantitative values for example of the carrier concentrations at which different behaviors occur will be somewhat lower than is realistic. The information available to us \cite{Gull10_clustercompare,Gull12b,Gull13,Gull13b} indicates that all of the qualitative features of the doping dependence are well reproduced by the $U=6t$ computations. As we will see, considerable insight can be obtained from examination of the $U$-dependence of the superconducting self energy in the special case $n=1$. The actual model also has an antiferromagnetic state, which competes with the superconducting state but is not studied here.

\begin{figure*}[tbh]
\includegraphics[width=0.95\textwidth]{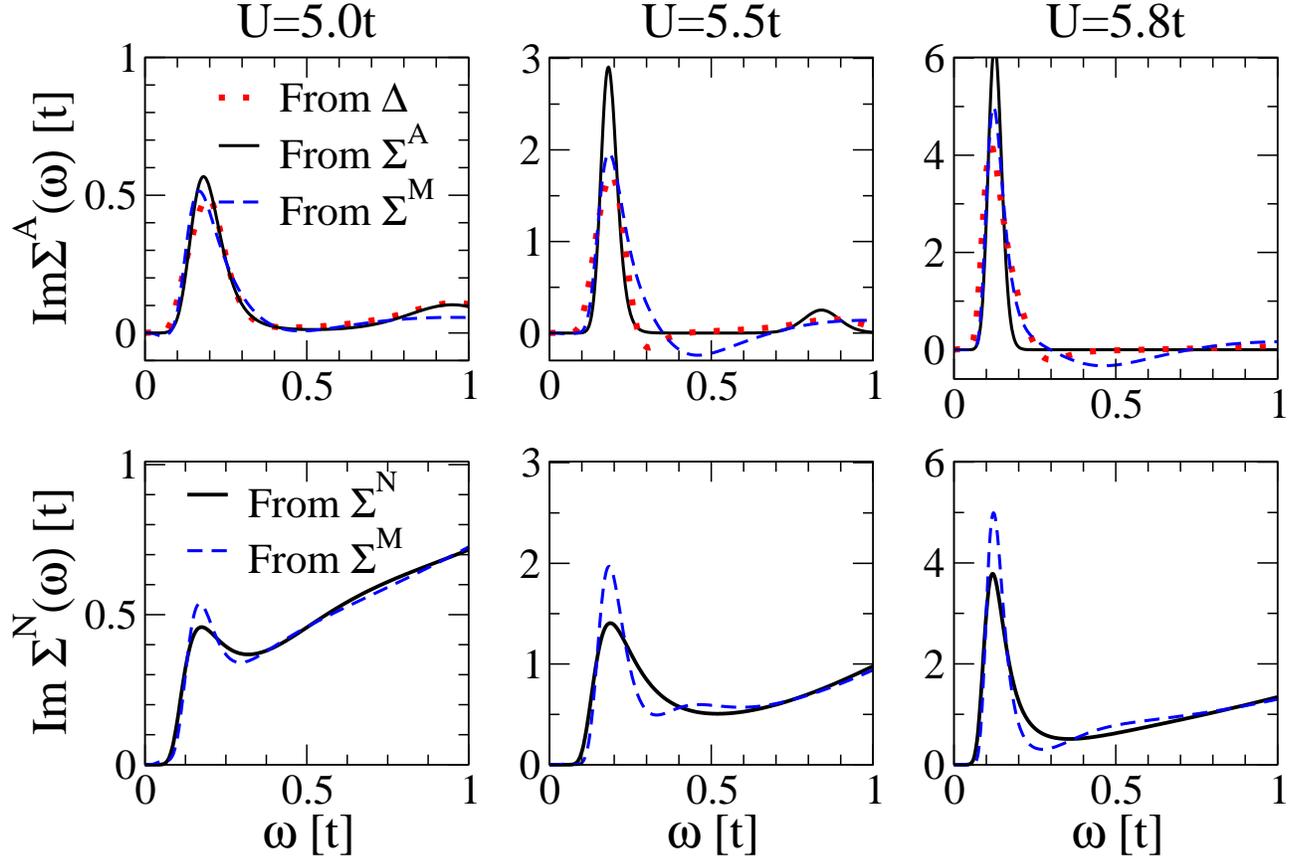}
\caption{Upper panels: imaginary part of anomalous component of self energy computed in the superconducting state at temperature $T=t/60$ and carrier concentration $n=1$ for the $3$ interaction values indicated, using three methods:  direct continuation of measured anomalous components of  self energy $\Sigma^A(i\omega_n)$ (solid lines, black online), reconstructed using Eq.~\ref{SigAfromSigpm} from a continuation of $\Sigma^M(i\omega_n)$ (dashed lines, blue online) and obtained from continuations of the gap function $\Delta(i\omega_n)$ and the normal component of the self energy (dotted lines, red online).  Lower panels: imaginary parts of normal component of self energy computed for the same parameters using direct continuation of measured normal components of self energy $\Sigma^N(i\omega_n)$ (solid lines, black online), and reconstructed using Eq.~\ref{SigNfromSigpm} from a continuation of $\Sigma^M(i\omega_n)$   (dashed lines, blue online).}
\label{fig:compareNandAnarrow}
\end{figure*}

\section{Analytical Continuation of Normal and Anomalous Self Energies \label{continuation}}
\subsection{Formalism}
The main objects of interest in this paper are the real-frequency normal (N) and anomalous (A) components of the  Nambu matrix electron self energy ${\mathbf \Sigma}(k,z)$ as well as the gap function $\Delta(z)$ defined from Eq.~\ref{Deltadef}. These are functions of a complex frequency argument $z$.  Our results for $\mathbf{\Sigma}$ and thus $\Delta$ are obtained on the imaginary frequency (Matsubara) points $z=i\omega_n=i(2n+1)\pi T$. Theorems from complex variable theory guarantee that knowledge of the function on the Matsubara points fully determines the function at all $z$, but direct inversion of Eq.~\ref{KKSigma} to obtain $\mathbf{\Sigma}^{(2)}(\omega)$ from measurements of $\mathbf{\Sigma}(i\omega_n)$ is a mathematically ill-posed problem. In this paper we invert Eq.~\ref{KKSigma} using the maximum entropy method \cite{Jarrell96} and verify the results with the Pad\'{e} method.\cite{Beach00}  

To continue the diagonal component $\text{Im}\Sigma^{N}(\omega)$ we  follow the methods employed for self energy continuation in Ref.~\onlinecite{Wang09a}.  While there are no rigorous methods to control errors in this procedure, our experience is that given sufficiently accurate input data (relative statistical errors smaller than $10^{-3}$ on each Matsubara point) this procedure produces reasonably reliable results for the lowest frequency features in the real-axis spectral function at low temperature and qualitatively reasonable results (estimates of characteristic energy scales and integrated areas) for the higher frequency features. The relative statistical errors in our calculations are typically smaller than $~10^{-4}$ so we have reasonable confidence in the qualitative features of the continuations.

A variation of the procedure outlined in \textcite{Wang09a} is needed to obtain the off-diagonal component of the self energy $\text{Im}\Sigma^{A}$ because   $\Sigma^A(i\omega_n)=\Sigma^{A}(-i\omega_n)$ so that the spectral function $\text{Im}\Sigma^{A}(\omega)=-\text{Im}\Sigma^{A}(-\omega)$ is an odd function of frequency and is thus not non-negative. We rewrite Eq.~\ref{KKSigma} as
\begin{align}
\Sigma^A(i\omega_n)&=\int \frac{dx}{\pi}\frac{x}{i\omega_n-x}\left(\frac{\text{Im}\Sigma^{A}(x)}{x}\right)
\nonumber \\
&=\Sigma^{(A)}(i\omega_n=0)+i\omega_n\int \frac{dx}{\pi}\frac{\frac{\text{Im}\Sigma^{A}(x)}{x}}{i\omega_n-x}
\end{align} 
and continue $\frac{\Sigma^A(i\omega_n)-\Sigma^A(0)}{i\omega_n}$ by standard methods.\cite{Gull14_glue} The normalization of  $\frac{\text{Im}\Sigma^{A}(x)}{x}$ is fixed from $\lim_{i\omega_n\rightarrow 0}\Sigma^A(i\omega_n)$: We obtain $\Sigma^{A}(i\omega_n=0)$ by fitting $\left(\Sigma^A\right)^{-1}$ at the three lowest positive Matsubara frequencies to a parabola. A similar procedure is used to continue $\Delta$.

\subsection{Non-negativity of $\text{Im}\Sigma^{A}(\omega)/\omega$}
While the usual maximum entropy analytic continuation formalism requires a non-negative spectral function,  there is no guarantee that $\text{Im}\Sigma^{A}(\omega)/\omega$ is of definite sign. For example, in  Migdal-Eliashberg theory the Coulomb pseudopotential leads to a negative contribution to  $\text{Im}\Sigma^{A}$ at frequencies of the order of the plasma frequency,\cite{Scalapino66} while a recent solution of the Eliashberg equations for a model involving two competing spin fluctuations also displayed a sign change in the gap function as frequency was increased above the lower of the two characteristic frequencies.\cite{Fernandes13}

For the two dimensional Hubbard model, our data are consistent with a positive definite $\text{Im}\Sigma^{A}(\omega)/\omega$ but we do not have a rigorous proof that this is always the case. Evidence for a positive definite $\text{Im}\Sigma^{A}(\omega)/\omega$ may be found from consideration of the particle-hole symmetric ($n=1$) situation. In this case  in the $(\pi,0)$ sector  the normal components of the hybridization function and  self energy are also particle-hole symmetric and are odd functions of Matsubara frequency, so that $\theta_\omega=\pi/2$ at all frequencies and there is a frequency-independent basis choice (the `Majorana combination' $c_\pm=\left(c^\dagger\pm c\right)/\sqrt{2}$) that diagonalizes the self energy matrix  in the $(\pi,0)$ sector. The corresponding `Majorana' self energy $\Sigma^M$ obeys the Kramers-Kronig relation
\begin{equation}
\Sigma^M(z)=\int \frac{dx}{\pi}\frac{s(x)}{z-x}.
\label{SigmaM}
\end{equation}
Thus analytical continuation of the Majorana combination of self energies gives direct access to $s$ from which the normal and anomalous components of the self energy can easily be reconstructed from Eqs.~\ref{spectral},\ref{Rdef} as

\begin{eqnarray}
\text{Im}\Sigma^N(\omega)&=&\frac{s(\omega)+s(-\omega)}{2},
\label{SigNfromSigpm}
\\
\text{Im}\Sigma^A(\omega)&=&\frac{s(\omega)-s(-\omega)}{2}.
\label{SigAfromSigpm}
\end{eqnarray}

Fig.~\ref{fig:compareNandAnarrow} presents a comparison of $\text{Im}\mathbf{\Sigma}$ obtained by direction continuation of the normal and anomalous parts of the Matsubara Green function and reconstructed using Eqs.~\ref{SigNfromSigpm} and \ref{SigAfromSigpm} from continuations obtained from the $(1,1)$ component of $\mathbf{\Sigma}$ in the Majorana basis (continuation of the $(2,2)$ component yields values differing by $\sim 10^{-3}$). A reconstruction of $\Sigma^A(\omega)$ from the continued $\Delta(\omega)$ and $\Sigma_o(\omega)$ using Eq.~\ref{Deltadef} is also shown.

We see that the methods yield very similar results with some differences in the height and width of the low frequency peak (the integrated peak areas, not shown, are the same). The differences between the curves are an indication of the continuation errors.  In the $\text{Im}\Sigma^A$ obtained from $s(\omega)$ a relatively small amplitude oscillation is present, which leads to a small negative value in some regions where the directly continued $\text{Im}\Sigma^A=0$ along with an overshoot (relative to the directly continued $\Sigma^A$) at slightly higher frequencies. Our experience is that these oscillations do not vary systematically with input data or details of the maximum entropy procedure, and we believe they  are artifacts of the continuation procedure related to the requirement that norm of the function is conserved in the entropy  minimization. We thus believe that $\text{Im} \Sigma^A(x)/x$ is generically non-negative for the $d_{x^2-y^2}$ superconducting state of the Hubbard model. This conclusion further supported by continuations (not shown) that we have performed using the Pad\'{e} method at both $n=1$ and $n\neq 1$. The Pad\'{e} method makes no assumptions as to the sign of the spectral function, and in the cases we have studied leads always  to a non-negative $\text{Im}\Sigma^{A}(x)/x$.  A $\Sigma^A(\omega)/\omega$ which was non-negative was also found by  \textcite{Civelli09b} in $N_c=4$ calculations using an exact diagonalization (ED) solver (we believe that the small negative excursions are artifacts of the ED method  but the issue deserves further investigation). 

\subsection{Pole structure and continuation of gap function}
Fig.~\ref{fig:compareNandAnarrow} reveals that in certain parameter regimes the normal and anomalous self energies exhibit   strong peaks at  relatively low frequencies. As will be discussed at length below, these peaks do not correspond to physical excitations of the system; rather they are the expression in the superconducting state of the physics of the normal-state pseudogap, which in the DCA  is associated with the formation of a low frequency pole in the self energy of the $(0,\pi)/(\pi,0)$ momentum sector.\cite{Lin10}  It is useful to discuss the pole structure in terms of the representation in Eq.~\ref{spectral}.  All of our data are consistent with the statement that $s(\omega)$ for the $(0,\pi)/(\pi,0)$ sector is the sum of a pole   and a regular part:
\begin{equation}
s(\omega)=D\delta\left(\omega-\omega^\star\right)+s_\text{reg}(\omega)
\label{polestructure}
\end{equation}
with $s_\text{reg}$  a smooth function of $\omega$. The pole structure was also noted in a very recent paper by Sakai and collaborators.\cite{Sakai14}

Retaining for the moment only the pole term and explicitly evaluating Eq.~\ref{spectral} and \ref{Rdef} using the Nambu angle $\theta^\star$ corresponding to $\omega^\star$ gives
\begin{eqnarray}
\mathbf{\Sigma}^{N}_\text{pole}(\omega)&=&D\frac{\omega +\omega^\star\cos\theta^\star}{\omega^2-\left(\omega^\star\right)^2},
\label{poleN}
\\
\mathbf{\Sigma}^{A}_\text{pole}(\omega)&=&D\frac{\omega^\star \sin\theta^\star}{\omega^2-\left(\omega^\star\right)^2}.
\label{poleA}
\end{eqnarray}

\begin{figure}[bth]
\includegraphics[width=0.9\columnwidth]{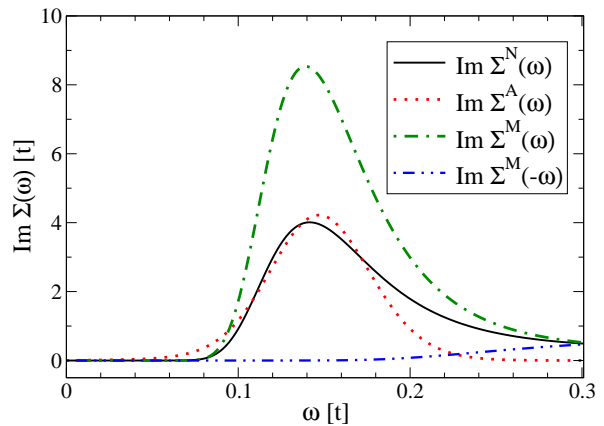}\caption{Imaginary part of normal ($\Sigma^N$) and anomalous ($\Sigma^A$) components of self energy and of $\Sigma^M$ at positive and negative frequency, computed by analytic continuation of Matsubara axis self energies for density $n=1$ and interaction $U=5.8t$ at temperature $T=t/60$.} 
\label{fig:selfenergiesU58}
\end{figure}

Thus in general we expect the normal and anomalous components of the self energy to have poles at exactly the same frequencies.  In the particle-hole symmetric case, where $\theta=\pi/2$, we expect the poles in the normal and anomalous parts of the self energy to have exactly the same amplitude. This is demonstrated in Fig.~\ref{fig:selfenergiesU58}, which shows the imaginary parts of the continuations of the normal and anomalous components of the self energy, along with the continuations of $s(\omega) = \text{Im}\Sigma^M(\omega)$ obtained from continuation of the $(1,1)$ component of the Majorana-basis representation of $\mathbf{\Sigma}(i\omega_n)$ and of $s(-\omega)$ obtained from continuation of the  $(2,2)$ component of the Majorana-basis representation of $\mathbf{\Sigma}(i\omega_n)$. That $s(-\omega)=0$ for $\omega$ in the vicinity of the pole is strong evidence of the cancellation.   In the general case both normal and anomalous components of the self energy have poles at $\pm \omega^\star$  but because $\cos \theta^\star\neq0$ the average of the  strengths of the poles in $\Sigma^N$  will  in general be greater than  the strengths of  the poles in $\Sigma^A$. However, as will be seen in our detailed examination of the self energy below, the background contributions are such that unambiguously identifying the pole strengths is not possible. 

It also follows from Eqs.~\ref{poleN}, \ref{poleA} and \ref{Deltadef} that the pole contribution to  the gap function $\Delta$ is
\begin{equation}
\Delta_\text{pole}=\frac{
D\sin\theta^\star\omega^\star
}
{
\omega^2-\left(
\left(\omega^\star\right)^2+D
\right)
}
\label{Deltapole}
\end{equation}
so that $\Delta$ does not have poles at $\omega=\pm\omega^\star$. All of our continuations are consistent with this result. 

The structure defined by Eq.~\ref{polestructure} creates challenges for analytical continuation.  Intrinsic errors in the analytical continuation process mean that independent continuations of the different components of $\mathbf{\Sigma}$ may lead to slightly different estimates of the pole positions, amplitudes and widths. This can be seen for example by comparison of the upper and lower panels of Fig.~\ref{fig:compareNandAnarrow} or by comparing $\Sigma^N$ and $\Sigma^A$ in Fig.~\ref{fig:selfenergiesU58}. The difficulties are exacerbated if particle-hole symmetry is broken because the continuation process may not place the  poles in $\Sigma^N$ at exactly opposite frequencies. We do not know how to control the continuation process so as to force the precise alignment of  the poles in the different components of $\Sigma$. For studies of the particle-hole symmetric situation, we work with $s$ defined from continuation of the $(1,1)$ component of $\mathbf{\Sigma}$ in the Majorana basis. In the doped case, we focus on the gap function.

\section{The normal and anomalous self energies, $(0,\pi)/(\pi,0)$ sector}\label{self_energies}
It has been accepted for many years that the two dimensional Hubbard model exhibits a Mott insulating phase at carrier concentration $n=1$ and large interaction, and a Fermi liquid phase at small interaction or carrier concentration sufficiently different from $1$. Work \cite{Huscroft01,Civelli05,Macridin06,Kyung06b,Park08placquette,Gull08,Liebsch08,Ferrero09,Werner098site,Gull09,Liebsch08,Liebsch09,Sakai09,Sakai10,Lin10,Gull10_clustercompare,Sordi10,Sordi11,Sordi12} has established that if superconductivity is neglected then the Mott and Fermi liquid phases are separated by a pseudogap regime, in which regions of momentum space near the zone face are gapped while regions of momentum space near the zone diagonal are not.  Mathematically, within DCA the gapping is a consequence of the appearance of a pole in the electron self energy pertaining to the $(0,\pi)$ sector.\cite{Werner098site,Gull09,Gull10_clustercompare} If the Mott insulator is approached by varying interaction strength in the particle-hole symmetric case ($n=1$, $\varepsilon_k=-\varepsilon_{k+(\pi,\pi)}$) the pole occurs at $\omega=0$;\cite{Werner098site} in a non-particle-hole symmetric situation (for example if the Mott insulator is approached by varying doping) the pole appears at a frequency close to, but slightly different, from zero.\cite{Gull09} In this section we show how the onset of superconductivity affects this pole structure.

\begin{figure}[t]
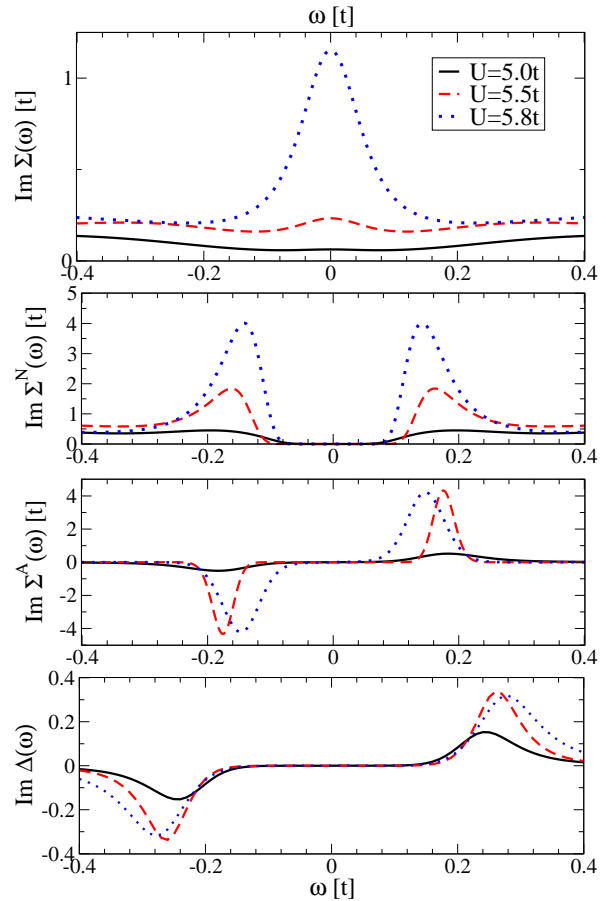

\includegraphics[width=0.9\columnwidth]{imnormselfofUbeta30}
\includegraphics[width=0.9\columnwidth]{imscselfNandAofUbeta60}
\includegraphics[width=0.9\columnwidth]{ImDeltaOmega_U_ajm}
\caption{Top panel: imaginary part of normal state self energy calculated for carrier concentration $n=1$ and $U$-values indicated, at temperature $T=t/30$.  Middle panels: imaginary part of normal  and anomalous components of self energy calculated in superconducting state at $T=t/60$. Lowest panel: imaginary part of gap function calculated in superconducting state at $T=t/60$. }
\label{fig:imNormSCselfofU}
\end{figure}

\begin{figure}[t]
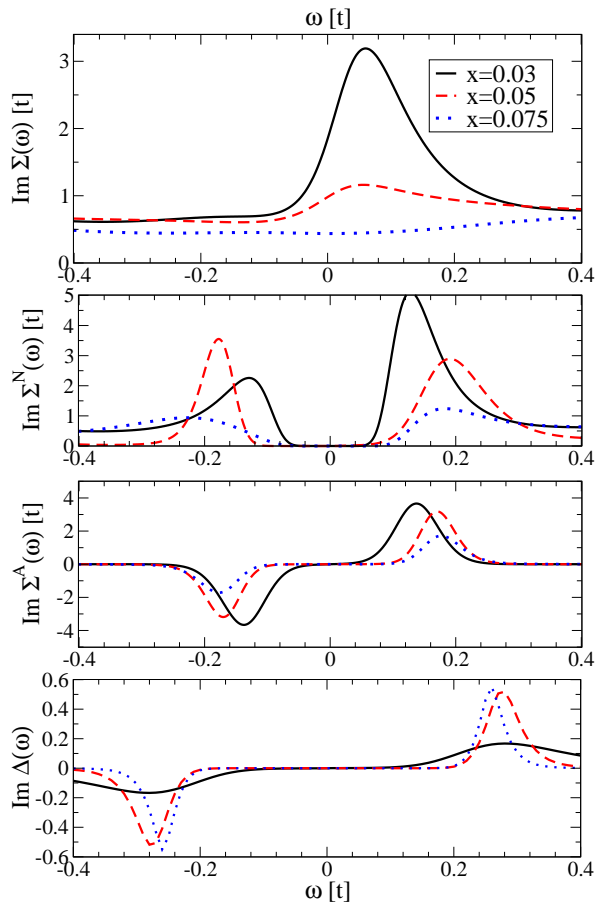

\includegraphics[width=0.9\columnwidth]{imnormselfofdeltaU6beta30}
\includegraphics[width=0.9\columnwidth]{imscselfNandAofdopingU6beta60}
\includegraphics[width=0.9\columnwidth]{ImDeltaOmega_doping_ajm}
\caption{Top panel: imaginary part of normal state self energy calculated for interaction strength $U=6t$ and dopings $x$ indicated, at temperature $T=t/30$.  Middle panels: imaginary part of normal and anomalous  components of self energy calculated in superconducting state at $T=t/60$. Lowest panel: imaginary part of gap function calculated in superconducting state at $T=t/60$. }
\label{fig:imNormSCselfofdoping}
\end{figure}

Fig.~\ref{fig:imNormSCselfofU} presents the imaginary parts of the analytically continued self energies and the gap function  for the three $U$-values at $n=1$ presented in Fig.~\ref{fig:compareNandAnarrow}. The upper panel shows results obtained in the normal state  at temperature $T=t/30>T_c$ (at $U=5.8t$ $T_c$ is slightly greater than $t/30$ but superconductivity has been suppressed here for clarity of presentation, {\it i.e.} $\Sigma^A$ set to zero).  At $U=5.0t$ the imaginary part has  the Fermi liquid form, with $\text{Im}\Sigma(\omega)$ exhibiting an approximately quadratic minimum at  zero.  As $U$ is increased a thermally broadened pole appears at $\omega=0$.  The pole increases rapidly  in strength as $U$ is increased. These results are consistent with our previous analysis of the pseudogap.\cite{Gull10_clustercompare,Lin10}  The middle two panels present the imaginary parts of the normal  and anomalous components of the self energy at a temperature $T=t/60<T_c$.  One sees that the pole centered at $\omega=0$   has split into two and appears with comparable strength in both the normal and anomalous components.  The pole frequency weakly decreases as $U$ is increased.  The larger width seen in the $U=5.8t$ calculation is likely to be an artifact of the analytical continuation. The lowest panel shows the imaginary part of the gap function. We see that the gap function is much smaller in magnitude than the self energy, that the  structure in the gap function occurs at a higher frequency than the structure in the self energy (as also noted by Sakai et al \cite{Sakai14}), and that the gap function varies less dramatically with interaction strength than does the self energy. 

Fig.~\ref{fig:imNormSCselfofdoping} shows the analogous plots as a function of doping at $U=6t$. The upper panel (normal state) shows again a pole that appears and grows in strength as doping is decreased. The particle-hole symmetry breaking provided by the doping means that the pole is slightly displaced from the origin. In the superconducting state (middle panels) the pole  splits; in $\Sigma^N$ the pole strength is different for the positive frequency than for the negative frequency pole. The strength of the pole in $\Sigma^A$ is of the order of that in $\Sigma^N$. The detailed line shapes of the poles depend to on the details of the continuation process. The difference in line shapes and small differences in pole position can lead to unphysical structures in calculated spectra. As in the interaction-driven case, there is no structure in $\Delta$ at the pole frequencies of $\Sigma$.  

Poles in the anomalous component of the self energy were reported by Civelli \cite{Civelli09b} and the alignment of poles in $\Sigma^N$ and $\Sigma^A$ and the cancellation of the $\Sigma$ poles in $\Delta$ were very recently discussed in the context of 4-site CDMFT calculations \cite{Sakai14}.

\section{The Superconducting Gap and the pseudogap \label{gap}}
In this section we present results for the energy gap in the superconducting and non-superconducting states. The gap may be estimated from an analytical continuation of the normal component of the Green function, but the inevitable broadening associated with continuations means that it is not clear a priori how to estimate a gap. As discussed in Ref.~\onlinecite{Wang09a} for Mott insulators, an alternative approach provides a more accurate estimate. This approach is based on the argument that for frequencies less than the gap, all imaginary parts are zero except for the poles in $\Sigma$, which do not correspond to physical excitations. Thus one may determine the  gap  energy $\omega_g$ from the vanishing of the real part of the denominator of the Green function. For the normal state, this criterion involves consideration of 
\begin{equation} 
\omega_g(k)=\varepsilon_k-\mu+\text{Re}\Sigma[\omega_g(k)].
\label{NSgapdef}
\end{equation}
For a given $k$ this equation will have two solutions, $\omega_g^\pm(k)$. The pseudogap is typically indirect (the $k$ that maximizes  $\omega_g^-$ is different from the $k$ that minimizes  $\omega_g^+$). The direct gap is determined as
\begin{equation}
\Delta_{PG}=\frac{1}{2}\min_k\left\{\omega_g^+(k)-\omega_g^-(k)\right\}.
\label{DeltaPGDef}
\end{equation}
The points of minimum direct gap are found to be the renormalized Fermi surface points $k$ satisfying $\varepsilon^\star_k=\varepsilon_k-\mu+\text{Re}\Sigma_e(\omega=0)=0$ and that at these points $\omega^+=-\omega^-=\Delta_{PG}$.

\begin{figure}[t]
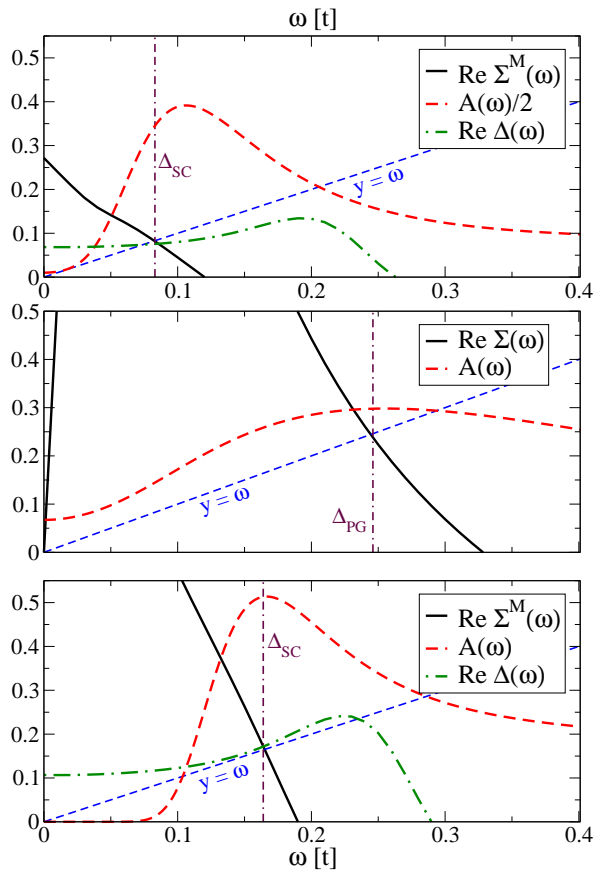

\includegraphics[width=0.9\columnwidth]{sigmaanddosU5beta60}
\includegraphics[width=0.9\columnwidth]{sigmaanddosU58beta30}
\includegraphics[width=0.9\columnwidth]{sigmaanddosU58beta60}
\caption{Determination of superconducting and pseudo gaps. Top panel:
The real part of the $\Sigma^M$ self energy (solid line, black online), the sector $(0,\pi)$ integrated spectral function $A(\omega)=\frac{1}{\pi} \text{Im} G(\omega)$ (dashed line, red online, scaled by $0.5$), and the real part of the gap function $\Delta(\omega)$ computed at $n=1$, $U=5$ and $T=t/60<T_c$. Middle  panel: Self-energy and spectral function in the normal state, for $n=1$, $U=5.8t$ and $T=t/30>T_c$. Bottom panel: same analysis as in top panel, for $n=1$, $U=5.8t$ and $T=t/60<T_c$.
Also shown are the line $y=\omega$ (blue online) and an estimate for the superconducting or pseudo gaps obtained from $\text{Re}\Sigma^M(\omega)=\omega$  (vertical line, purple online).}
\label{fig:gaptest}
\end{figure}

In the superconducting state, these ideas lead to the consideration of $\det\left[\mathbf{G}^{-1}(k,\omega)\right]$. Rearranging the expression for $\mathbf{G}^{-1}$ gives
\begin{eqnarray}
\det\left[\mathbf{G}^{-1}(k,\omega)\right]&=&\left(1-\frac{\Sigma_o(\omega)}{\omega}\right)^2\times
\label{Ginv}\\
\bigg(\omega^2&-&\varepsilon_k^\star(\omega)^2-\Delta^2(\omega)\bigg)
\nonumber
\end{eqnarray}
with $\Delta$ defined in Eq.~\ref{Deltadef} and
\begin{equation}
\varepsilon_k^\star(\omega)=\frac{\varepsilon_k-\mu+\Sigma^N_e(\omega)}{1-\frac{\Sigma_o(\omega)}{\omega}}.
\label{estardef}
\end{equation}
We then  define the gap frequency  $\omega_g(k)$ in the superconducting state as the frequency at which the real part of $G^{-1}$ vanishes, i.e. as 
\begin{equation}
\omega_g(k)=\pm\sqrt{\text{Re}[\varepsilon^\star_k(\omega_g(k))]^2+\text{Re}[\Delta(\omega_g(k))]^2}.
\label{scgap}
\end{equation}
We define the Fermi surface as the locus of $k$-points for which $\varepsilon^\star_k(\omega=0)=0$. We will find that the the $\omega$ dependence of $\varepsilon_k^\star$ is modest so that we may identify the gap $\Delta^{SC}$  in the superconducting state as the value of $\omega_g$ which solves  $\omega_g=\text{Re}\left(\Delta(\omega_g)\right)$.

In the particle-hole symmetric case where $\Sigma_e^N=0$  we may alternatively write the gap equation as the solution of \begin{align}\omega_g=\text{Re}\Sigma^M(\omega_g).\label{gapeqphsym}\end{align}

\begin{figure}[t]
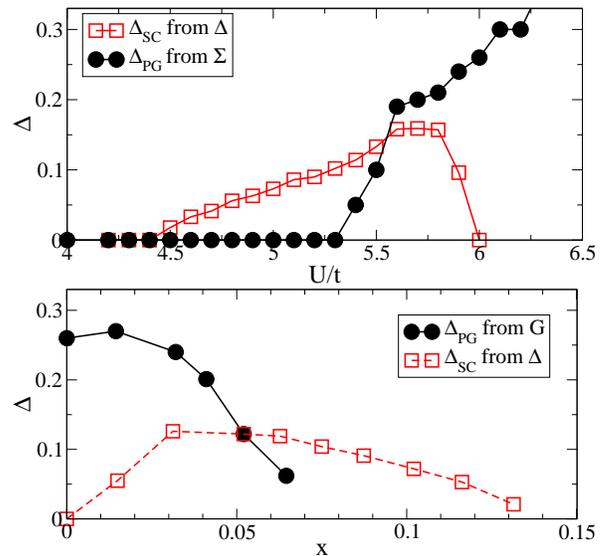

\includegraphics[width=0.9\columnwidth]{pgandscgapofUn1}
\includegraphics[width=0.9\columnwidth]{pgandscgapofdopingu6}
\caption{Energy of lowest excitation in superconducting state (temperature $T=t/60$, dashed lines, red online) and normal state (temperature $T=t/60$ but superconductivity suppressed, solid line, black online)  computed for different interaction strengths at carrier concentration $n=1$ (upper panel) and different carrier concentrations at interaction strength $U=6t$ (lower panel)  as described in the text.}
\label{fig:gapofUanddelta}
\end{figure}

Fig.~\ref{fig:gaptest} demonstrates this procedure. The upper panel shows results obtained in the superconducting state at $U=5$ (the normal state is not shown because at this $U$ there is no normal state pseudogap).  The figure plots the spectral function (imaginary part of continued Green function), the real part of the Majorana combination of the self energy and the real part of the gap function as a function of frequency. We see by comparing the continued spectral function to the self energy and gap curves  that the criteria $\omega=\text{Re}\Delta(\omega)$ or $\omega=\text{Re}\Sigma^M(\omega)$ identifies a point close to that at which the spectral function is maximal. We believe that the appearance of weight at lower frequencies in the spectral function comes from artificial broadening induced by the continuation process. 

The middle panel shows the same analysis in the normal state at $U=5.8t$.  The normal state pseudogap is evident, and again we see that the quasiparticle equation picks out as the gap the point at which the spectral function is maximal. The lower panel shows the superconducting state, also at $U=5.8t$.  We also see that at this $U$ value the line $y=\omega$ is  tangent to (in fact very slightly below) the $\text{Re} \Delta(\omega)$ curve at the point that one would naturally identify as the gap.  That $\text{Re}\Sigma^M(\omega)$ intersects $y=\omega$ while $A(\omega)$ is peaked at the point of near tangency suggests that the absence of an exact intersection is a continuation artifact and that if all quantities were exactly and consistently continued the line $y=\omega$ would intersect the $\text{Re}\Delta(\omega)$ line. 

Comparison of the middle and lower panels of  Fig.~\ref{fig:gaptest} shows that the superconducting gap is unambiguously smaller than the normal state pseudogap, consistent with results presented in Ref.~\onlinecite{Gull13}. Also, comparison of the lower panel of Fig.~\ref{fig:gaptest} to Fig.~\ref{fig:compareNandAnarrow} shows that for this value of $U$ the pole in the self energy lies inside the excitation gap of the superconductor, further confirming that the self energy pole does not represent a physical excitation of the system. 

\begin{figure}[t]
\includegraphics[width=0.9\columnwidth]{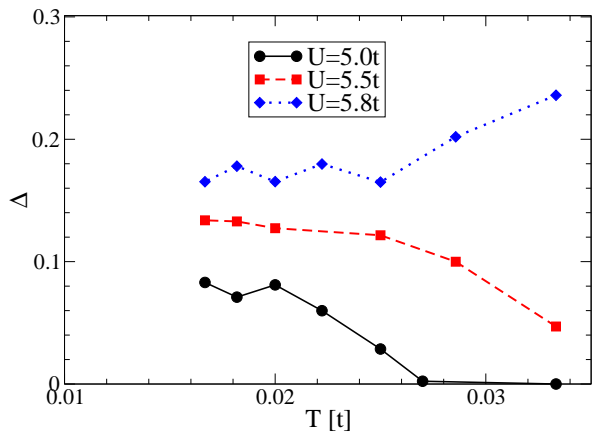}
\caption{Temperature dependence of gap size computed as discussed in text for carrier concentration $n=1$ and interaction strengths indicated
}
\label{fig:Tdepofgap}
\end{figure}

The two panels of Fig.~\ref{fig:gapofUanddelta} show the dependence of the low-T ($T=t/60\sim 60K$) gap  on interaction strength at $n=1$, computed from Eq.~\ref{gapeqphsym} and the dependence of the low-T gap carrier concentration at $U=6t$, computed from Eq.~\ref{scgap}. In the weak interaction or large doping limits, there is no superconductivity. As the doping decreases or interaction strength increases the gap increases smoothly down to  the low doping/high interaction end of the phase diagram (note that at the two endpoints of the superconducting phase, $(n=1,U=5.9)$ and $(U=6,x=0.02)$ the superconducting state is not fully formed so the gap values are not meaningful). Also shown on the plots is the normal state pseudogap, computed at the same low temperature  $T=t/60$ by suppressing superconductivity in the calculation. One sees that in the low doping/strong interaction regime, turning on superconductivity leads to a decrease in the energy of the lowest-lying excitation. 

Finally, Fig.~\ref{fig:Tdepofgap} shows the temperature dependence of the energy gap at the three $U$ values considered above. We see that in the moderate coupling regime, there is no normal state gap and the superconducting   gap increases from zero as the temperature is decreased through the transition temperature. At intermediate coupling a small but non-zero gap is already present in the normal state and the gap increases with the onset of superconductivity, while at stronger coupling the gap actually decreases as the temperature is decreased into the superconducting state. 

\section{Fermi surface spectra\label{spectraFS}}
In this section we present spectra computed at the Fermi surface using continued self energies. The issues discussed above relating to the difficulties of analytical continuation in non-particle-hole symmetric case mean that we limit ourselves here to study of the interaction-driven transition at density $n=1$ but all of the information available to us suggests that the behavior in the doped case shares the essential features found at half-filling. 

\begin{figure}[b]
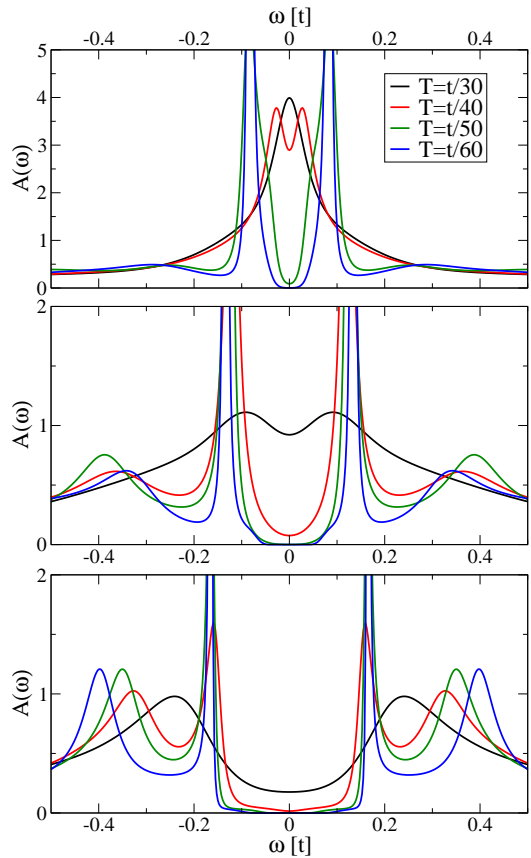

\includegraphics[width=0.8\columnwidth]{U50tempdep}
\includegraphics[width=0.8\columnwidth]{U55tempdep}
\includegraphics[width=0.8\columnwidth]{U58tempdep}
\caption{Spectral functions computed at carrier density $n=1$ for momenta on the Fermi surface in the $(0,\pi)$ sector for $U=5.0$ (top panel) $5.5$ (middle panel) and $5.8$ (lower panel) at temperatures indicated.}
\label{fig:Tdepspectra}
\end{figure}

Fig.~\ref{fig:Tdepspectra} shows the temperature evolution of the electron spectral function computed for a momentum on the Fermi surface at the antinode ($(\pi,0)$) at carrier density $n=1$ and different interaction strengths. The upper panel shows results obtained for a moderate interaction $U=5t$. We see that the normal state spectral function is peaked at the chemical potential $\omega=0$. The onset of the superconductivity  ($T_c$ is between $t/30$ and $t/40$) induces a suppression of the low frequency density of states. The gap grows and sharpens as temperature is decreased. For the lower temperatures one sees that the spectral function has a sharp quasiparticle peak, a weak minimum at slightly higher frequencies, followed by a weak maximum at yet higher frequencies.  This structure in the spectral function is frequently observed in photoemission \cite{Dessau91,Ding96,Norman97,Campuzano99,Kordyuk02,Kordyuk06,Carbotte11} and scanning tunneling microscopy \cite{Coffey93}  experiments on copper-oxide high-$T_c$ materials and is referred to as a ``peak-dip-hump'' feature. The higher frequency ``hump'' is typically interpreted as arising from the interaction of electrons with some kind of bosonic excitation, with the hump frequency determined by the boson energy and the minimum energy to create an electron-hole pair. 

The middle panel of Fig.~\ref{fig:Tdepspectra} shows shows results obtained for  an intermediate interaction $U=5.5$. We see that a weak minimum is evident in the density of states even  temperature $T=t/30>T_c$; this weak suppression of the density of states marks the onset of the  `pseudogap'. As the temperature is decreased below the transition temperature the low frequency intensity drops  very rapidly. The gap (peak in the spectral function) increase and saturates at a value rather greater than that found in the moderate interaction case. The peak-dip-hump structure is more evident.  In this case the ``hump'' energy increases as T decreases except for the lowest temperature.  

The lower panel of Fig.~\ref{fig:Tdepspectra}  shows results obtained for the strongest interaction ($U=5.8$). In this case the normal state pseudogap is well established even at $T=t/30$.  The transition to superconductivity is associated with the formation of a quasiparticle peak which lies inside the pseudogap and with a shift outwards of the second peak structure. For both $U=5.5$ and $U=5.8$ the energy of the second peak  increases as temperature decreases, with the exception of $U=5.5$, $T=t/60$. We believe that at this particular $U$ and $T$ the continuations are of lower quality than at other parameter values. For example the errors associated with back continuation are slightly larger. 

\begin{figure}[t]
\includegraphics[width=0.95\columnwidth]{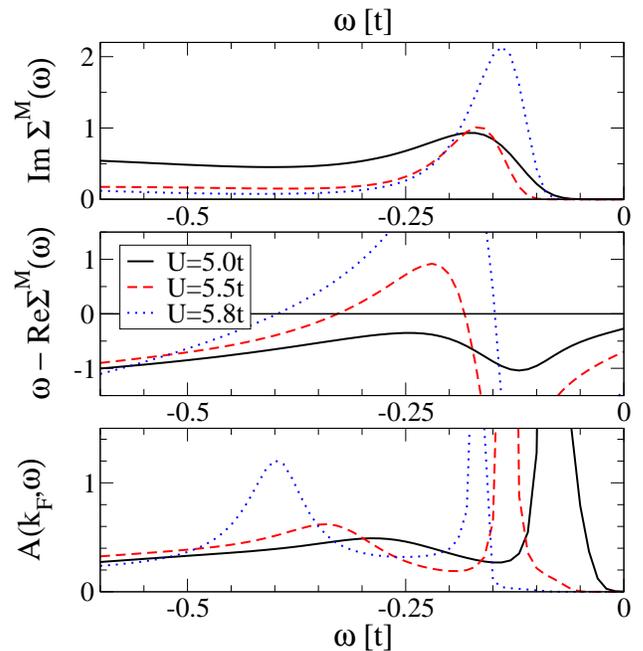}
\caption{Comparison of real and imaginary parts (middle and top panel) of the $\Sigma^M$ Majorana component of self energy to Fermi surface spectral function (lower panel) for carrier concentration $n=1$, temperature $T=t/60$ and $U=5$ (solid lines, black online), $U=5.5$ (dashed lines, red online) and $U=5.8$ (dotted lines, blue online). In the middle panel the real part of the self energy is presented as $\omega-\text{Re}\Sigma(\omega)$.}
\label{fig:peakdiphump}
\end{figure}

The  question of the origin of the ``peak-dip--hump'' structure  has been discussed extensively in the literature on angle-resolved photoemission in the cuprates.\cite{Grabowski96,Norman97,Chubukov98,Abanov99,Abanov00c,Kordyuk02,Borisenko03,Carbotte11,He11} The most widely accepted explanation is that the higher energy ``hump'' is evidence for a ``shakeoff'' process in which an electron emits a bosonic  excitation such as a spin fluctuation \cite{Chubukov98,Abanov99,Abanov00c,Borisenko03} or a phonon,\cite{Lanzara01} while the peak feature is the quasiparticle excitation above the gap. In this picture  the onset of the  hump feature determined as the sum of the energy of the boson and the superconducting gap energy, (structure in the bare dispersion associated with interbilayer hopping may also play a role in certain materials). \cite{Kordyuk02,Borisenko03} Such a shakeoff process would appear in the imaginary part of the self energy as an upward step, corresponding to the opening of a scattering channel.  However, a quantitative and generally accepted identification of the boson responsible for the hump energy  is lacking.

Mathematically, structures in the spectral function may be understood from the fundamental expression
\begin{equation}
A(k,\omega)\sim \frac{\text{Im}\Sigma(\omega)}{\left(\omega+\mu-\varepsilon_k-\text{Re}\Sigma(\omega)\right)^2+\left(\text{Im}\Sigma\right)^2}
\label{Astructure}
\end{equation}
where for clarity we have not explicitly written the change of basis matrices $\mathbf{R}$ (Eq.~\ref{Rdef}).  From Eq.~\ref{Astructure} we see that structure can arise from resonance ($\omega+\mu-\varepsilon_k-\text{Re}\Sigma(\omega)\approx 0$) or, off resonance, from an increase in $\text{Im}\Sigma$. The former case produces the quasiparticle peak of Fermi liquid theory; the latter is the origin of the  ``shakeoff'' explanation of the peak-dip-hump structure.\cite{Norman97,Chubukov98,Abanov99,Abanov00c,Kordyuk02,Carbotte11}  Fig.~\ref{fig:peakdiphump} investigates the origin of the ``hump'' structure in the present calculation by comparing the computed  spectral function (lower panel, evaluated at the Fermi surface $\varepsilon_k=\mu$) to the electron self energy (heavy lines, upper panel) in the Majorana basis that diagonalizes the Nambu Greens function at all $\omega$ for $n=1$. The real part is presented in the combination $\omega-\text{Re}\Sigma(\omega)$. We see that the location of the ``hump'' feature in fact does not correspond to any significant feature in the imaginary part of the self energy (top panel), so that it cannot be interpreted as an onset of scattering. Instead, inspection of $\omega-\text{Re}\Sigma(\omega)$ energy (middle panel) shows that for the two stronger couplings the frequency of the hump  corresponds to the frequency at which the real part of the inverse Greens function vanishes, while for the weaker coupling the hump frequency corresponds to a point where the absolute value of $\omega-\text{Re}\Sigma(\omega) $ is minimized.  In other words, the ``hump'' is a resonance phenomenon, not a scattering phenomenon, and its frequency does not correspond directly to the energy of  any excitation of the system.

\begin{figure}[t]
\begin{overpic}[scale=0.45,]{U5.eps}
\put(5,60){\includegraphics[width=0.2\columnwidth]{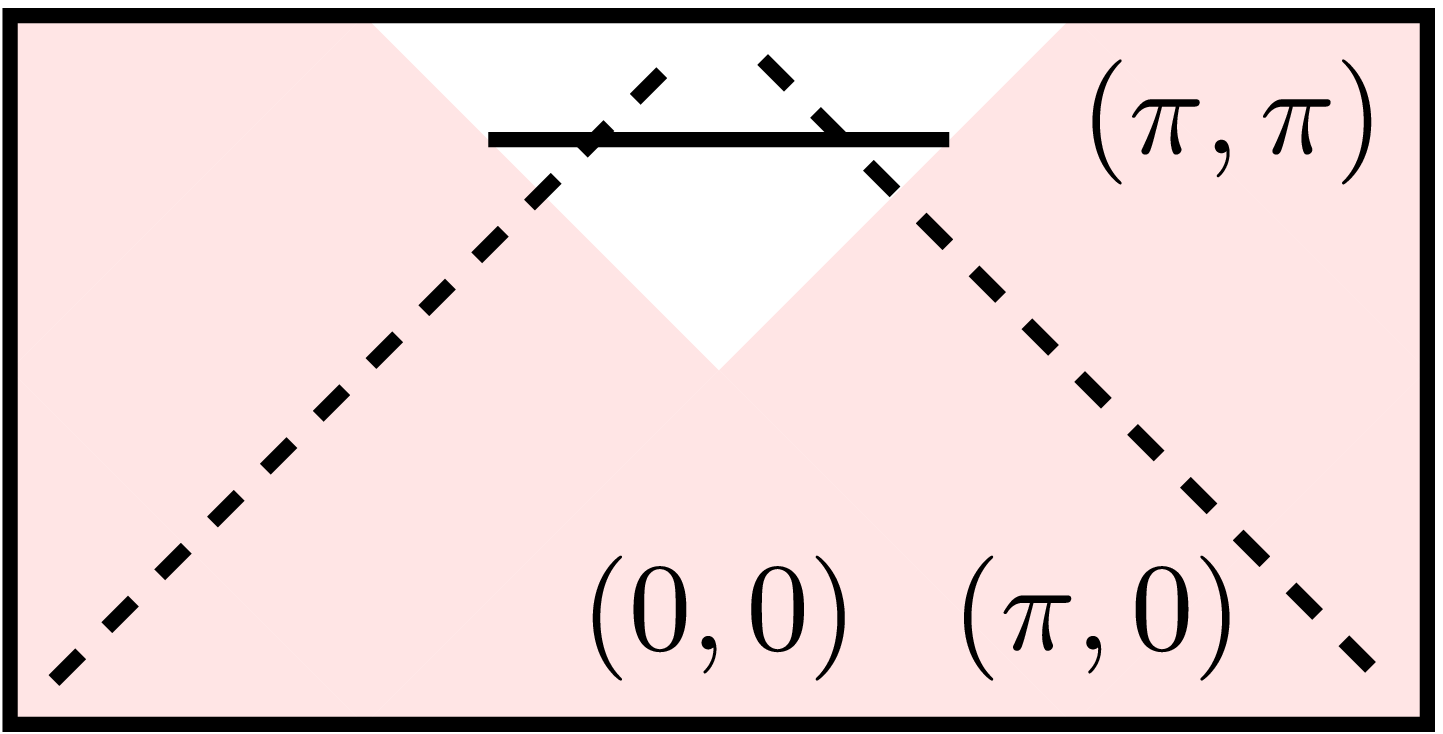}}
\end{overpic}
\caption{Energy dependence of electron spectra calculated at density n=1 and interaction $U=5$ for momenta $-0.325\pi\leq k_x\leq0.325\pi$ and $k_y=0.825\pi$. Inset: Illustration of the momentum cut (horizontal line) through the noninteracting Fermi surface (dashed line); see also Fig.~\ref{fig:pdtiling}. Left panel: normal state ($T=t/30$), right panel: superconducting state ($T=t/60$). Dot-dashed curves (red online) correspond to momenta inside the Fermi surface and solid curves (black online) to momenta outside the Fermi surface. The dashed curves (blue online) indicate traces with the momenta set equal to the Fermi momentum. 
}
\label{fig:U5kdependence}
\end{figure}

\section{Momentum-dependent Spectra \label{spectra}}
In this section we present and discuss the momentum dependence of  spectral functions calculated from the normal and anomalous components of the self energy.  The left panel of Fig.~\ref{fig:U5kdependence} shows a sequence of energy distribution curves (EDC) calculated in the normal state for a sequence of momenta cutting across the Fermi surface in the $(\pi,0)$ sector for the moderate interaction strength $U=5.0$. A quasiparticle peak is visible, which disperses through the Fermi surface. The right  panel shows EDC at the same momenta, this time in the superconducting state. Comparison of the two panels reveals the behavior expected of a moderate-coupling BCS-like superconductor, with the gap having the greatest effect on the Fermi surface trace and the superconductivity-induced  particle-hole mixing producing a peak dispersing away from zero as momentum is increased above the Fermi level. 

\begin{figure}[t]
\includegraphics[width=0.8\columnwidth]{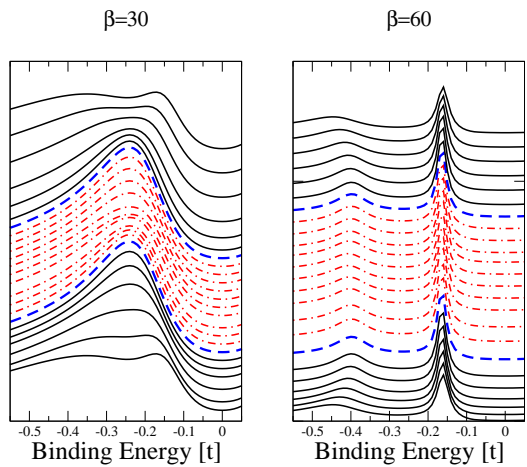}
\caption{Energy dependence of electron spectra calculated at density n=1 and interaction $U=5.8$ for momenta $-0.325\pi\leq k_x\leq0.325\pi$ and $k_y=0.825\pi$. Left panel: normal state ($T=t/30$), right panel: superconducting state ($T=t/60$). Dot-dashed curves (red online) correspond to momenta inside the Fermi surface and solid curves (black online) to momenta outside the Fermi surface. The dashed curves (blue online) indicate traces with the momenta set equal to the Fermi momentum. 
}
\label{fig:U58kdependence}
\end{figure}

Fig.~\ref{fig:U58kdependence} shows the EDC curves obtained for a strong interaction $U=5.8$. The weakly dispersing and rather broadened  normal  state pseudogap is seen in the left panel. The right panel shows that  the onset of superconductivity produces a very weakly dispersing peak at an energy well inside of the pseudogap. This essentially non-dispersing zone-edge feature is characteristically observed in photoemission experiments on high $T_c$ copper oxide superconductors.\cite{Norman97,Shen98,Lanzara01,He11} In the present calculation the weak dispersion arises mathematically from the very strong frequency dependence of the self energy caused by the proximity of the pseudogap pole (cf Eq. ~\ref{Astructure}).

\section{Superconducting effects on zone diagonal spectra\label{diagonal}}
\begin{figure}[t]
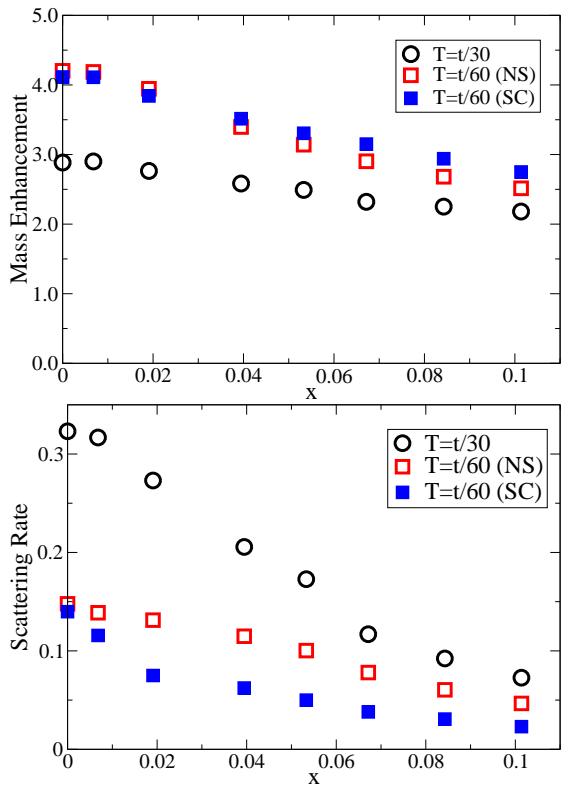

\includegraphics[width=0.85\columnwidth]{sectorBmass}
\includegraphics[width=0.85\columnwidth]{sectorBscatt}
\caption{Dependence on doping of zone diagonal mass enhancement $\frac{m^\star}{m}$ (upper panel) and Fermi surface scattering rate $\Gamma_0$ (lower panel) extracted from Matsubara-axis data (Eq.~\ref{SigmasectorB} ) for temperatures  indicated in normal and superconducting state. }
\label{fig:sectorBmassandscatt}
\end{figure}

In the cuprates and in the Hubbard model the dramatic effects of superconductivity are visible in the electronic states near the zone face ($(0,\pi)/(\pi,0)$) point, because it is at this point that the superconducting gap and the pseudogap are maximal.  But in addition to opening, or changing the value of, a gap,  the onset of superconductivity may affect other aspects of the physics, for example by changing the density of final states that enter into scattering processes \cite{Dahm05} or, more profoundly, by changing the electronic state itself and thus the nature of the scattering mechanisms. Some understanding of these effects may be gleaned from consideration of electronic states with momentum along the zone diagonal. The superconducting gap vanishes for these momenta and thus any effects on electron propagation must arise from changes in scattering and electronic state. 

We find that at all dopings and interaction strengths we have studied the electronic self energy in the zone diagonal momentum sector has approximately the Fermi liquid form. In particular the self energy does not have a low frequency pole. Its imaginary part is minimal at zero frequency and the value at zero frequency, $\Gamma_0$,  decreases as temperature $T\rightarrow 0$.  The real part is linear in frequency over a reasonable range of  low frequencies, allowing us to define a zone diagonal mass enhancement $\frac{m^\star}{m}=1-\partial \Sigma/\partial\omega$ from the Matsurbara data (note also that in DCA, the self-energy is k-independent except at the sector boundaries, so the k-derivative contribution to the disperson renormalization is not relevant). We extract estimates of the Fermi level scattering rate $\Gamma_0$ and mass enhancement $\frac{m^\star}{m}$ by fitting the lowest four Matsubara frequencies  to the cubic form
\begin{equation}
\text{Im}~ \Sigma (i\omega_n)=-\Gamma_0\text{sign}(\omega_n)-\left(\frac{m^\star}{m}-1\right)i\omega_n+C\omega_n^3...
\label{SigmasectorB}
\end{equation}

Representative results are shown in Fig.~\ref{fig:sectorBmassandscatt}.  We see that the mass enhancement has a modest doping dependence, and decreases markedly as temperature is decreased. Superconductivity has only a small effect on the mass enhancement. At low doping the onset of superconductivity leads to a very small decrease in the mass enhancement; at higher doping the effect is of opposite sign and is slightly larger. It is interesting to note that  the change in sign of the superconducting contribution to the mass enhancement occurs at a lower doping than the change from potential energy-driven to kinetic energy driven pairing discussed in Ref.~\onlinecite{Gull12b}. Thus the change in electronic state associated with the onset of superconductivity has a weaker effect on the nodal quasiparticles than it does on other properties. 

The normal state scattering rate also has a dramatic doping dependence, especially at the higher temperature, and at all dopings exhibits a marked temperature dependence. The effect of superconductivity on the scattering rate is much more noticeable than the effect on the mass enhancement: the onset of superconductivity leads to an almost factor of two drop in the Fermi surface scattering rate, except at the very lowest doping which is right on the boundary of superconducting phase. These finding are consistent with angle-resolved photoemission measurements on Bi$_2$Sr$_2$CaCu$_2$O$_{8-\delta}$.\cite{Valla06} Fig.~4c of Ref.~\onlinecite{Valla06} reveals that the onset of superconductivity leads to an approximately factor of two decrease in the MDC width (a good proxy for scattering rate) below the extrapolation of the normal state rate to low temperature, while Fig.~3 of Ref.~\onlinecite{Valla06} reveals a much smaller change in the mass enhancement (albeit of opposite sign to that predicted here).  The more coherent nature of the zone diagonal quasiparticles is also qualitatively consistent with conclusions drawn from pump-probe experiments,\cite{Graf11} but because our calculations are restricted to equilibrium a direct comparison cannot be made. 
 
\section{Conclusions \label{Summary}}
Two characteristic features of the high transition temperature copper-oxide superconductors are superconductivity with  $d_{x^2-y^2}$ symmetry and a `pseudogap'\cite{Huefner08}, a suppression of electronic density of states for momenta near the Brillouin zone face ($(0,\pi)$ point). The interplay between these two phenomena  has been the focus of considerable attention in the literature. In this paper we present cluster dynamical mean field calculations of the electronic self energy and spectral function of the normal and superconducting phase of the two dimensional Hubbard model that shed new light on the subject. 

The correspondence of the results to the essential features of superconductivity in the cuprates is striking. We find, consistent with a large body of experimental literature, that when superconductivity emerges from the pseudogap regime the onset of superconductivity is associated with the appearance of very weakly dispersing states {\em inside} the pseudogap and with the appearance of a peak-dip-hump structure in the spectral function. Superconductivity affects the zone-diagonal states via a significant ($\sim$ factor of 2) decrease in the scattering rate. We have also determined the superconducting and pseudogaps accurately and have shown that the superconducting gap systematically increases as doping is decreased or interaction strength is increased, until it  abruptly drops to zero at the low-doping/high interaction boundary of the superconducting phase. These results support the notion that the pseudogap and superconductivity are competing phenomena and strengthen the case made  in prior papers \cite{Gull12b,Gull13,Gull13b,Gull14_glue} that as P. W. Anderson predicted in 1987 \cite{Anderson87} the Hubbard model contains the essential physics of the high-$T_c$ observed in layered copper-oxide materials.

Mathematically, we find (in agreement with previous work \cite{Werner098site,Gull10_clustercompare,Lin10,Gull13}) that the pseudogap is a Mott-transition-like phenomenon (``sector-selective Mott transition'') associated with the appearance of a low frequency pole in the self energy in the zone-face self energy.  The interplay between superconductivity and the pseudogap is controlled by the superconductivity-induced changes in the pole structure.   In particular,  in our calculation the `hump' feature in the spectral function arises mathematically from a zero-crossing in the real part of the self energy and not from the onset of a scattering process, in other words, not from a bosonic excitation at all.   On the technical side we have clarified the  mathematical structure of the pseudogap-induced  poles associated with the normal and anomalous self energies, shown how the inevitable analytical continuation errors make it difficult to construct real-frequency spectra  in situations where the  pseudogap poles are important.  The pole structure has also been discussed by \textcite{Sakai14} who present an interesting alternative interpretation.

It is important to consider the limitations of the methods used here. The essential technical step is the  use of continuous-time auxiliary field methods \cite{Gull08} and submatrix updates. \cite{Gull10_submatrix} These methods enable  highly accurate simulations at temperatures as low (in physical units) as $60K$, far below the basic scales of the model and, crucially, well below the superconducting transition temperature.  However, even with these improvements, obtaining  results of the needed precision at the required low temperatures requires approximations. 

The key approximation used in this paper is the cluster dynamical mean field approximation.\cite{Maier06} In the `DCA' form used here \cite{Hettler98} this amounts to approximating the normal and anomalous components of the electron self energy as piecewise constant functions of momentum, taking different values in each of $N_c$ momentum sectors that tile the Brillouin zone. We have adopted the $N=8$ approximation. This is the smallest momentum decomposition that permits a clear separation between the zone face and zone diagonal regions of momentum space. Previous work \cite{Maier08,Gull10_clustercompare,Gull13} indicates that although we do not have quantitative convergence to the $N=\infty$ limit, the $N=8$ approximation correctly captures the physics of the normal state pseudogap. 

A second approximation is the use of an interaction $U\leq 6t$ which is likely to be slightly weaker than needed to quantitatively capture the physics of the cuprates. This approximation is needed because computation time increases rapidly as $U$ increases (as $U^3$, with additional complications from the sign problem) and we needed to undertake a broad survey of parameter space.  For similar computational reasons we restricted attention to the particle-hole symmetric version of the model (second neighbor  hopping $t^\prime$ and further neighbor hoppings set to zero).  

Finally, because the basic computations use imaginary time methods, obtaining spectra of interest requires an analytical continuation method. Analytical continuation is an ill-posed problem.  We used maximum entropy analytical continuation methods (which are widely employed but  essentially uncontrolled) to extract real frequency information.  This requires extremely high quality Monte Carlo data, further constraining the parameter ranges that could be examined. In this context the properties of the anomalous self energy are of particular importance. With the exception of the Pad\'{e} method, all methods known to us require a positive definite spectral function. While we have presented evidence that in at least some cases the spectral function associated with the anomalous self energy has an appropriate positivity property, the  possibility of a sign change at high frequency cannot be ruled out. At high frequencies the anomalous part is very small and the intrinsic limitations of the continuation process mean that small systematic errors could induce (or mask) a sign change. 

A key physical  limitation of our work is that we have not considered other ordered states (for example N\'{e}el antiferromagnetic or  striped order) that might preempt the phases considered here. The dynamical mean field method captures (within the rather coarse momentum resolution of the cluster dynamical mean field method) fluctuations associated with these states, but because we have symmetrized over spin degrees of freedom, long range ordered antiferromagnetic states are excluded. Also our calculation lacks the momentum resolution needed to provide a clear account of striped states.   Thus we view the results as providing a reasonable qualitative account of the properties of the superconducting phase and pseudogap regime of the two dimensional Hubbard model, but not as a quantitatively accurate account of the properties of the Hubbard model. 

For these reasons, extensions of the work presented here would be desirable. Pushing the calculation on large clusters to somewhat larger $U$ so that the $n=1$ endpoint is well within the Mott insulating phase should become feasible as computer power improves. Study of larger $U$ would provide more insight into the interplay of superconductivity and Mott physics. Extending the calculations to $N=16$ site approximation would similarly be  useful. In particular, examination of differences between gap values and spectra calculated with $N=8$ and $N=16$ will provide insight into the quantitative aspects of the results.    If feasible, real-time  calculations and comprehensive Pad\'{e} continuations, that could investigate the possibility of sign changes in the anomalous component of the self energy would be reassuring. On the conceptual side, a deeper understanding of the pseudogap state and of the meaning of the self energy poles would be desirable. Most importantly, investigation of competing states such as N\'{e}el antiferromagnet and stripe orders is needed. 

{\it Acknowledgements}: We thank M. Civelli, T. Maier, D. Scalapino, M. Norman, and A.-M.~S. Tremblay for helpful discussions.   The research was supported by NSF-DMR-1308236 (A.J.M.) and the Sloan foundation (E.G.). A portion of this research was conducted at the National Energy Research Scientific Computing Center (DE-AC02-05CH11231), which is supported by the Office of Science of the U.S. Department of Energy. Our continuous-time quantum Monte Carlo and Maximum entropy codes are based on the ALPS\cite{ALPS20,ALPS_DMFT} libraries.

\bibliography{refs_shortened}
\end{document}